\documentclass[aps,prx,amsmath,amssymb,twocolumn,superscriptaddress]{revtex4-2}
\usepackage{graphicx}
\usepackage{physics}
\usepackage{mathtools}
\usepackage[ruled,vlined]{algorithm2e}
\usepackage{booktabs}
\usepackage{isomath}
\newcommand{\pd}[2]{\frac{\partial #1}{\partial #2}}
\begin{document}

\title{Reinforcement-learning based matterwave interferometer in a shaken optical lattice}

\author{Liang-Ying Chih}
\author{Murray Holland}
\affiliation{JILA, NIST, and Department of Physics, University of Colorado, 440 UCB, Boulder, CO 80309, USA}

\date{\today}

\begin{abstract}
We demonstrate the design of a matterwave interferometer to measure acceleration in one dimension with high precision. The system we base this on consists of ultracold atoms in an optical lattice potential created by interfering laser beams. Our approach uses reinforcement learning, a branch of machine learning, that generates the protocols needed to realize lattice-based analogs of  optical components including a beam splitter, a mirror, and a recombiner. The performance of these components is evaluated by comparison with their optical analogs. The interferometer's sensitivity to acceleration is quantitatively evaluated using a Bayesian statistical approach. We find the sensitivity to surpass that of standard Bragg interferometry, demonstrating the future potential for this design methodology.
\end{abstract}

%\keywords{}

\maketitle

\section{Introduction}
Quantum metrology is an important field of quantum physics with the goal to make accurate and precise measurements of important physical quantities, ideally at a level that surpasses that achievable by any classical approach. 
In particular, extensive efforts have been made to develop metrological devices based on interference and detection of either electromagnetic waves or matterwaves. By exploiting the quantum aspects of superposition and entanglement, one can potentially achieve high sensitivity to phase shifts, and this has inspired a wide variety of applications, including detecting gravitational waves \cite{LIGO}, measuring the fine-structure constant \cite{FineStructure}, testing the universality of free fall~\cite{EquivalencePrinciple}, and inertial sensing for GPS navigation \cite{Gyroscope,Navigation}.

The usual direct-design approaches to engineering complex quantum systems that consist of many degrees of freedom or many particles are typically founded on experience with simpler analogs or intuition for underlying mechanisms. This naturally leads to a paradigm that is most accessible in terms of understanding, but incorporates human bias that may potentially generate non-optimal solutions. With this perspective in mind, we point out there are a few purely systematic methods that are often used as a way to develop unbiased strategies, including optimal control \cite{optimalControl} and optimization algorithms such as the Nelder-Mead simplex~\cite{Nelder-Mead} or simulated annealing \cite{simulatedAnnealing}. Utilizing these methods can allow one to explore solutions with more complicated forms with the potential to reach closer-to-optimum control protocols.

Recently, it has been shown that quantum design may benefit tremendously from a branch of machine learning based on trial and error, known as reinforcement learning, that aims to employ machines to find the optimal strategy for accomplishing a specific task.  One of the reasons for its success is that the learning framework is decoupled from human intuition, and therefore may explore novel solutions that have been previously undiscovered. Moreover, in situations where problems are so complex or extensive that na\"ive brute-force algorithms are considered unfeasible, such as playing games like {\em go}\/ and {\em chess}~\cite{alphaGo, alphaZero1, alphaZero2}, sophisticated reinforcement-learning algorithms have been developed to enable the machines to perform at a level that exceeds human capability. There are many systems in quantum physics that fit naturally within this scope due to their underlying complexity. For example, reinforcement learning has been applied to control and study phase transitions of many-body quantum systems \cite{RL_manyBody}, to find strategies for quantum error correction \cite{RL_QEC1, RL_QEC2}, to design quantum circuits \cite{RL_qCircuit}, to prepare novel quantum states \cite{RL_KapitzaOscillator, Mackeprang2020, RL_Bell}, to find protocols for quantum communication \cite{RL_qCommunication}, and to improve quantum sensors~\cite{RL_qSensor}.

\begin{figure*}
\begin{center}
\includegraphics[width=\textwidth]{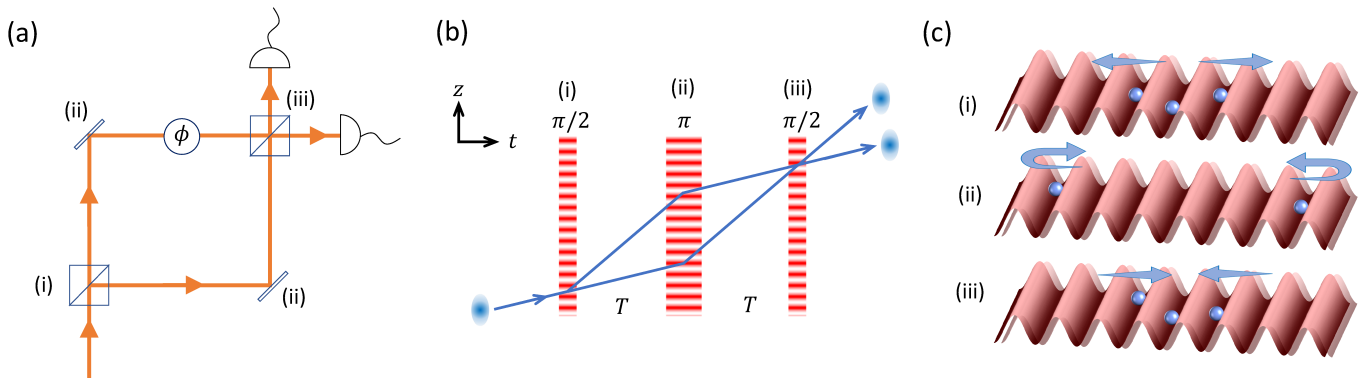}
\end{center}
\caption{Interferometers are composed of (i) a beam splitter, (ii) mirrors, and (iii) a recombiner. Examples shown are; (a) an optical Mach-Zehnder interferometer (using half-silvered or conventional mirrors), (b) a Bragg interferometer (using three short light pulses of varying pulse area, $\pi/2$ or $\pi$), and (c) a shaken lattice interferometer. In (c), the shaken lattice mimics the interferometer components by splitting, reflecting, and recombining the atoms, through design of a specific shaking function for each case.}
\label{fig:Intro}
\end{figure*}
In this paper, we will investigate the potential for reinforcement learning to improve upon exisiting matterwave interferometery. 
In Sec.~\ref{sec:model}, we describe the physical model for the interferometric system. We outline our reinforcement learning approach in Sec.~\ref{sec:RL} and explain the key concepts crucial to understanding its application. In Sec.~\ref{sec:components}, we demonstrate how the reinforcement-learning framework may be used to design a beam splitter and a mirror, and present the protocols that are learned. Following this we show how the components can be cascaded together to form a complete interferometer. In Sec.~\ref{sec:analysis}, we evaluate the resulting performance of the interferometer through a Bayesian statistical analysis.

\section{Physical Model} \label{sec:model}
Although the topology of interferometers can vary somewhat, the archetypal design can be thought of as the Mach-Zehnder interferometer \cite{Mach, Zehnder}. This device is sensitive to the differential phase accumulated between two alternate paths [see Fig.~\ref{fig:Intro}(a)]. 
Mach-Zehnder interferometers are composed from three essential components; a beam splitter that separates the wave coherently into two directions, mirrors that reflect both parts, and a recombiner where the waves are brought back together and constructively or destructively interfere. While in the optical case, components that split or reflect light beams are readily available (e.g., half-silvered or conventional mirrors), for matterwave optics, the components have to be generated through the careful control of laser-atom interactions. The traditional way to do this is through Bragg diffraction [see Fig.~\ref{fig:Intro}(b)]~\cite{atomInterferometry1, atomInterferometry2}, in which a sequence of three short light pulses are used to separate, reflect and recombine the matterwave.  

An alternative approach is possible for atoms moving in an optical lattice potential where the intensity pattern can be shifted backward and forward in time~\cite{Weidner1, Weidner2, Weidner3}. The elementary components, i.e., a beam splitter, reflectors, and a recombiner, in that case may be implemented by `shaking' the lattice in a tailored pattern. If we denote the canonical position and momentum operators of an atom by $\hat{x}$ and $\hat{p}$, respectively, with $m$ the atom mass, the optical lattice system can be described by the Hamiltonian
\begin{equation}
    \hat{H}(t) = \frac{\hat{p}^2}{2m} - \frac{V_0}{2}\cos\bigl[2k_L\hat{x} + \phi(t)\bigr]\,,
    \label{eq:Hamiltonian}
\end{equation}
where $k_L$ is the laser wavenumber, and $\phi(t)$ is the time-dependent phase difference between two counter-propagating lasers. The lattice, with constant amplitude~$V_0$, is shaken through the variation of $\phi(t)$, since that parameter determines the position of the nodes and the anti-nodes of the standing wave intensity.
%At this point, we neglect atom-atom interactions since the atom number is small. Therefore, the single-atom Hamiltonian is sufficient to describe the system. 

Unlike a typical Mach-Zehnder interferometer, which is formally a two-port device and transforms quantum states according to simple SU(2) group rotations, the relevant eigenstates of the shaken lattice potential consist of many Bloch states that can be coupled by the time-dependence of the laser phase, $\phi(t)$. In some sense this situation represents a highly multi-path form of interferometry since the Bloch basis provides many accessible paths for the quantum wave function to explore. While this establishes a rich evolution, and could provide metrological benefit, it makes it more difficult to design an intuitive control protocol. It is for this goal of obtaining a high performance solution among a complex landscape that we are led to consider reinforcement learning as a design approach.

\section{Model-free reinforcement learning for design} \label{sec:RL}
In order to understand our design philosophy, we first need to address the important considerations to make when using machines to find solutions to design tasks. In this section we will present the principal ideas and concepts that will form the underlying foundations of our learning-based methodology. Since there are a variety of methods available, we begin with a discussion of the main structure that we will employ.

\begin{figure}
\centering
\includegraphics[width=\columnwidth]{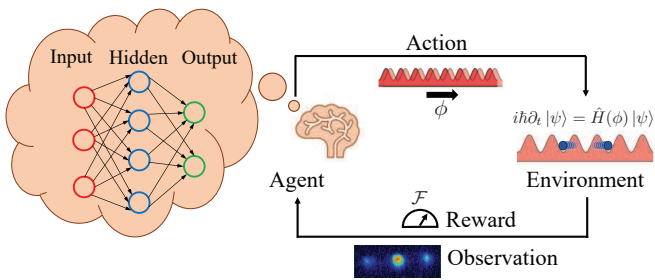}
\caption{Framework of reinforcement learning. The agent chooses an action, the environment responds to the action, and gives the next state and a reward as feedback. In our design task, the action is the translation of the optical lattice, the environment evolves according to the Schr\"odinger equation, the observation is the momentum population distribution, and the reward is a function of the quantum fidelity. Illustrated on the left is an example neural network as the decision-making agent. The neural network takes the state represented on its input layer, passes it through a hidden layer, and generates a vector of Q-values at its output layer.}

\label{fig:RL1}
\end{figure}

Reinforcement learning consists of a closed loop in which an {\em agent} invokes {\em actions}\/ based on the observed {\em state} of an {\em environment}, and an environment provides {\em rewards}\/ to the agent based on the observed outcome of its actions.
The agent is tasked with the goal to discover the sequence of actions for which it receives the highest possible terminal reward~\cite{RLSutton} (see Fig.~\ref{fig:RL1}). It does this by trial and error, iteratively improving its actions in such a way as to corral the environment towards a target configuration.

We are primarily interested in a specific kind of reinforcement learning, known as model-free learning~\cite{RLSutton}, where the agent has no detailed insight as to the structure of the environment, and cannot know {\em a~priori}\/ what effect an action will have on the environment state. Here the agent learns and makes decisions even when the environmental model may not be fully known. This general setting can even be applied to the situation in which the environment is represented by an experimental apparatus that cannot be fully understood, and the reward is derived from experimental observation and feedback. Using this approach significantly expands the potential scope of our work, since it is precisely this aspect that distinguishes model-free reinforcement learning from classical optimal control methods. In classical optimal control, the optimization depends heavily on the mathematical form of the model and, for complex systems, can be difficult to implement in practice.

Typically, reinforcement learning problems are Markov decision processes, where the probability of transition to the next state is only dependent on the current state and the current action, and not on prior history. Evolution in this framework is referred to as a {\em trajectory}, where the state is initially prepared, and then a sequence of actions and corresponding state updates are performed. The trajectory steps continue until the chain comes to an end when a predetermined condition is reached, which might be a certain number of steps, or a terminating state of the environment. An important concept is the idea of an {\em optimum trajectory} starting from any state, which is a trajectory such that the terminal reward is the maximum possible. 
We use what is known as Q-learning to make the action decisions, as we now describe. Although this approach is not the only possible choice, it is appropriate when the effect of an action on the environment state is deterministic, which will be the case here.

Since there is potentially an enormous dimensionality of the state space and the action space, a brute-force search is not possible, and we employ a neural network to learn and to approximately optimize the crucial decision-making task. That is, we define the input nodes of a neural network to be a representation of an arbitrary vector in the state space. We ascribe to the output of the neural network a vector of quality factors, simplified to Q-values, $Q(s,a)\in\mathbb{R}$, where each element represents the desirability for a possible action, $a$, given an input state,~$s$. Ideally  the neural network should take {\em any}\/ arbitrary state as input and output a distribution in which the maximum Q-value corresponds to the best next action to perform.

We train the neural network in the following manner. The agent will utilize a policy in order to decide on the next action for a given state, for example, by taking the action associated with the maximum Q-value from the neural network output vector. The environment, in some initial state $s$, receives the action $a$ from the agent, and then updates the state, symbolically denoted as
$$
s\xrightarrow{a}s'\,.
$$
The environment then reports back to the agent the reward, $r(s')$. As a step in a sequence, this leads to the important concept of the discounted cumulative reward, known simply as the {\em return}. The return, $Y(s')$, is defined as the combination of a current reward, $r(s')$, and every future reward that would be generated in an optimum trajectory seeded by~$s'$. Since any current action will have decreasing influence on future decision making as the steps become more distant, it is useful to deweight each consecutive reward by a discount factor, $\gamma\in[0,1]$, i.e.,
\begin{equation}
    Y(s')\equiv r(s')+\gamma r(s'')+\gamma^2 r(s''')+\ldots
    \label{eq:series}
\end{equation}
where
$$
s'\xrightarrow{a'} s''\xrightarrow{a''} s'''
\xrightarrow{a'''}\ldots
$$ 
are steps along an optimum trajectory.

The learning process is now framed as the minimization of the squared difference between the return, $Y(s')$, and the network output, $Q(s,a)$, where the minimization is accomplished by varying the internal weights and biases that constitute the neural network. Should the learning be perfected, the squared difference would be zero, implying that the Q-value and corresponding return would be identical. We emphasize that, while mathematically complete, this learning definition is a formal construction that is intractable as written, since if we knew the optimum trajectory there would be no need to carry out the learning at all. In fact, the only information that is accessible about the future component of $Y(s')$ is the inference one can make from the neural network output itself, introducing a self-consistency element to the optimization problem. In other words, the series in Eq.~(\ref{eq:series}) is approximated as
\begin{equation}
    r(s'')+\gamma r(s''')+\ldots\approx\max_{a'}\bigl[Q(s',a')\bigr]
    \label{eq:future}
\end{equation}
so that
\begin{equation}
    Y(s')=r(s')+\gamma \max_{a'}\bigl[Q(s',a')\bigr].
    \label{eq:cumulative}
\end{equation}
The final result given in Eq.~(\ref{eq:cumulative}) encapsulates all the principal concepts that establish a closed learning formulation that can be computationally implemented.

We employ several technical but important improvements that increase the convergence, such as to replace the neural network used for the calculation of the future return, as approximated in Eq.~(\ref{eq:future}), with a second identical neural network, the target network. The target network is updated at each step from the Q-network by a weighted average leading to increased stability~\cite{DQNNature, doubleDQN}. Another useful trick is to employ an $\epsilon$-greedy policy in the training stage~\cite{RLSutton}, which simply means that the action associated with the highest Q-network output value is not always chosen, but with $\epsilon$ likelihood, a random action is selected. This allows the neural network to explore a more extensive state space and to avoid becoming trapped in a locally optimal solution. Finally, as is common to many forms of machine learning, a variety of batch sampling methods, in this case for a replay buffer of previously experienced transitions, $(s, a, s')$, can be used to assist in the efficiency and stability of the learning minimization stage~\cite{Lin1992}. For further details of the algorithmic structure we refer the reader to Appendix~\ref{app:train}.

\section{Component design}
\label{sec:components}
We now use the Q-learning framework to design the essential elements for building an interferometer, i.e. a splitter, reflectors, and a recombiner. One aspect that is worth emphasizing is that the standard Mach-Zehnder interferometer topology may not always be the optimal layout for determining an accumulated quantum phase. However, for simplicity, and even though it constrains our solution space, we will limit the discussion here to the design of only the individual interferometer components. This will allow us to evaluate the efficacy of the learned protocols by direct comparison with the device characteristics of the analogous Bragg interferometer. In the future, it may prove beneficial to allow for more freedom in the design topology, and to thereby ascertain whether this leads to further performance gains.

We have developed the learning terminology around words such as `state' and `environment' where they have been given a specific meaning. We are now going to apply the methodology to the design task of controlling a complex quantum system. Unfortunately we will encounter the issue that terms such as state and environment have conflicting and completely different meanings in the theory of open quantum systems. In particular, to enable an efficient machine learning algorithm with an addressable number of input nodes, the state space that we use is typically a compact representation of a quantum state or operator. The environment, which denotes the system that we wish to control and that is evolving under our imposed actions, is a completely different use of the term  to that common in open quantum system literature. In the next design descriptions, we will provide the dictionary for mapping the learning terminology to the quantum variables. 

\subsection{Beam splitter}

The beam splitter is a target-state design problem that can be specified as follows. We initialize a quantum state in the ground state of an optical lattice potential, where the lattice depth $V_0$ is set to $10~E_r$ ($E_r=\hbar k_L^2/2m$ is the recoil energy). The task is to shake the lattice in such a way as to split the atoms from the ground state into an approximate equal superposition of $\ket{\pm p_0}$ momentum states. We will focus on $p_0=4\hbar k_L$ here, which we find to balance well the trade-off between large momentum splitting and high frequency components necessary for the shaking function. If the target quantum state is a lattice eigenstate it is a stationary state under free evolution, avoiding the need for a free evolution protocol. As a consequence, we assign the target quantum state to be the third-excited Bloch state with zero quasimomentum. This eigenstate has the attractive property that it closely approximates the desired superposition of momenta, although not perfectly, and can be strongly coupled to the ground state by a reasonably simple shaking solution.

The state space for learning, i.e., the vector space containing $s$, is a compact representation of the Hilbert space. For this learning task, we define $s$ to be a vector of populations (i.e., probability distribution) for momenta chosen from a comb of discrete values, $\{2n\hbar k\}$ for $n\in\mathbb{Z}$, $|n|<n_{\text{max}}$. This choice is motivated by the fact that the stimulated absorption and emission of photons only couple momenta separated by this quantized spacing. The possible actions, $a$, are also reduced to a discrete set of possibilities, each action corresponding to a specific constant phase to be applied during the corresponding step (see Appendix~\ref{app:RL_split}). This means that with these associations, the deep Q-network is specified as mapping momentum populations at its input layer to Q-values at its output layer associated with a finite set of possible lattice phases to be applied during the next update. 

In order to apply the action generated by a choice of lattice phase and to thereby determine the updated momentum distribution that results, we need to specify the environmental model. In principle, this could be experimental or theoretical, and even if theoretical, could include stochastic effects that arise from dissipation and noise. However, for simplicity, we will take the most basic formulation. This means that our environmental model will update a pure quantum state through the Schr\"odinger equation evolution for an isolated system. The reward will be based on the fidelity between the evolved quantum state and the target quantum state, i.e., the third-excited Bloch state, as calculated from the modulus-square inner product.

Actually, this environmental model allows us to accomplish two design tasks at the same time. Once we find a protocol for splitting, the recombining task can be achieved by simply implementing the time-reversed protocol, a consequence of the time-reversal symmetry underpinning the Schr\"odinger evolution and the time reversal symmetry of both the initial and target states. 
 
\begin{figure}[ht]
\centering
\includegraphics[width=\columnwidth]{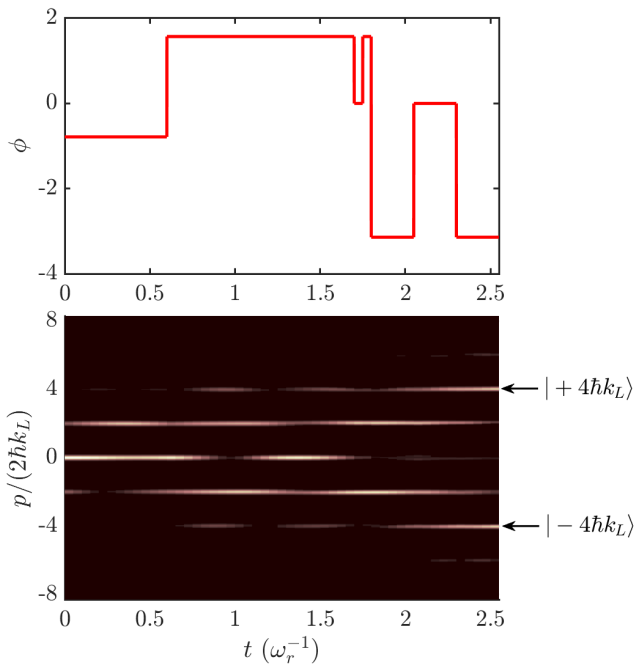}
\caption{Shaking function for splitting. The learned lattice phase, $\phi(t)$, (top) is allowed to take on one of a discrete set of 5 possible values that span the range shown. The momentum probability distribution (bottom) is initialized to the ground state of the lattice, and at the terminal time well-approximates the desired superposition for a beam splitter.}
\label{fig:splitting}
\end{figure}

A reinforcement learning outcome for the beam splitter is shown in Fig.~\ref{fig:splitting}, which illustrates the learned shaking function, $\phi(t)$, and the corresponding evolution for the momentum probability distribution that results. Even though this sequence of phases does not appear to have a predictable structure, it is apparent that at the terminal time the momentum distribution has the anticipated form of two well-defined peaks at $\pm4\hbar k_L$. The calculated fidelity of the state is approximately 95\%. It would be possible to optimize further by expanding the set of possible phase elements, by increasing the total evolution time, or by optimizing the learning cycles. However, any real experiment may possess imperfections and other aspects that are not well described by our isolated system's Schr\"odinger evolution, and the level of performance of our design is sufficient for the task at hand.  

\subsection{Mirror} 
 The second task is to reflect the momentum from $\ket{\pm p_0}$ to $\ket{\mp p_0}$, that is, corresponding to a matterwave mirror. We should point out at the start that this task is essentially different in character from the beam splitter, because it is not a target-state but a target-operator design problem. The mirror is defined by the desired map 
 $$\alpha\ket{p_0}+\beta\ket{-p_0}\rightarrow\alpha\ket{-p_0}+\beta\ket{p_0}
 $$
 for any arbitrary $\alpha$ and $\beta$. This implies that the target unitary is any operator $\hat{U}_{\text{target}}$ that satisfies 
 \begin{equation}
     \hat{P}\hat{U}_{\text{target}}\hat{P}=\ket{p_0}\bra{-p_0}+\ket{-p_0}\bra{p_0}
     \label{eq:mirror}
 \end{equation}
where $\hat{P}=\ket{p_0}\bra{p_0}+\ket{-p_0}\bra{-p_0}$ is the projector onto the 2-dimensional subspace. The design goal is therefore to shake the lattice as a function of time, such that the corresponding unitary evolution, $\hat{U}(t)$, approximates any $\hat{U}_{\text{target}}$ that satisfies Eq.~(\ref{eq:mirror}) at the terminal time,
\begin{figure}[ht]
\centering
\includegraphics[width=\columnwidth]{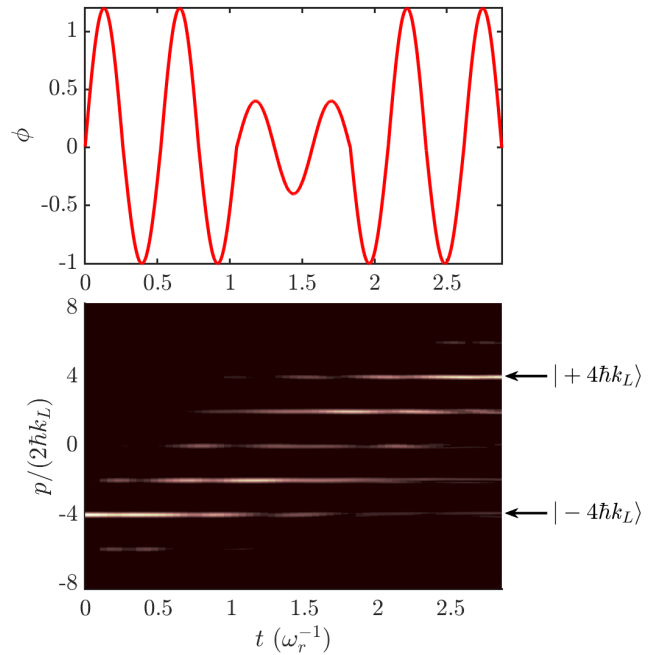}
\caption{Shaking function for reflection. The learned lattice phase, $\phi(t)$, (top) is sinusoidal and the amplitude is allowed to take on any one of a discrete set of 5 values at each half-cycle. The momentum distribution (bottom) is prepared in $-4\hbar k_L$ and well-approximates $4\hbar k_L$ at the terminal time.}
\label{fig:mirror}
\end{figure}
The vector space containing $s$ is defined as a compact representation, in this case, of the set of quantum unitary operators. Given the mirror behavior, a suitable choice for $s$ for this learning task is a vector of norms of selected matrix elements of the unitary represented in momentum space, i.e., $|\bra{\pm p_0}\hat{U}(t)\ket{p=2n\hbar k}|^2$ for $n\in\mathbb{Z}$, $|n|<n_{\text {max}}$. For the set of possible actions, $a$, we use a different parameterization of the phase than for the beam splitter design. We found that the quality of the mirror is very sensitive to the frequency of the shaking function. The characteristic frequency, $12 \omega_r$, where $\omega_r=\hbar k_L^2/2m$ is known as the recoil frequency of the lattice, corresponds to the kinetic energy difference between $\ket{4\hbar k_L}$ and $\ket{2\hbar k_L}$, and works especially well. Consequently, we enforce a sinusoidal $\phi(t)$ at frequency $12\omega_r$, and at each half-cycle, the agent chooses the action $a$ as the amplitude of the sinusoidal shaking function selected from a small set of possible values (see Appendix~\ref{app:RL_mirror}). This means here that the deep Q-network is specified as mapping a compact representation formed from the norm of a subset of elements of the unitary matrix to Q-values associated with the amplitude of the sinusoidal shaking function for its next half cycle.

We also need to define what is meant by the environment in this case. The environmental model establishes the unitary operator update during a shaking function half cycle. If, as before, we take the environmental model to be a closed quantum system, the evolution is given by solving the Schr\"odinger equation of motion for the time evolution operator
\begin{equation}
    i\hbar\frac{d\hat{U}(t)}{dt}=\hat{H}(t)\hat{U}(t)\,.
\end{equation}
The reward is associated with the channel fidelity in the $d_s=2$ dimensional subspace spanned by $\ket{\pm p_0}$~(see \cite{channelFidelity})
\begin{equation}
  \mathcal{F} = \frac{1}{d_s(d_s+1)} \left[\Tr(M M^\dagger) + |\Tr(M)|^2\right]  
  \label{eq:channel-fidelity}
\end{equation}
where the density operator is $M = PU_\mathrm{target}^{\dag}U(T)P$. 

An example outcome from reinforcement learning for the mirror is shown in Fig.~\ref{fig:mirror}. This illustrates the learned shaking function and corresponding momentum probability density that this generates starting from~$-4\hbar k_L$. From matrix elements of the resulting unitary operator, we also verify that the phase relation between the coefficients of the $\pm 4\hbar k_L$ states is preserved. As for the beam splitter case, it would be difficult to anticipate the form of $\phi(t)$, and yet the resulting quantum state evolution closely approximates that expected for a mirror. The channel fidelity reaches 93\% at the terminal time, which as discussed earlier could still be improved upon, but nevertheless is sufficient for our purpose.

\subsection{Matterwave interferometer} 
\begin{figure}
\centering
\includegraphics[width=\columnwidth]{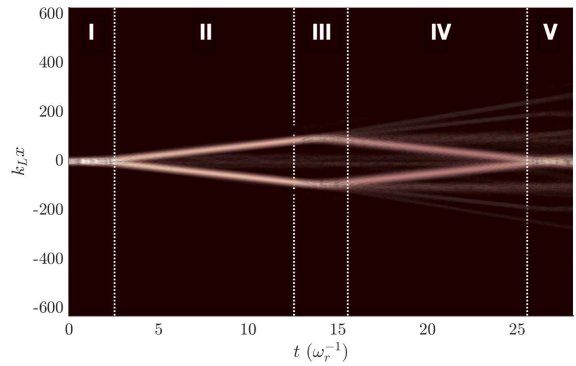}
\caption{Time evolution of the matterwave density throughout the entire interferometry sequence. The white dotted lines separate the plot into five regions. In region I, we apply the splitting protocol. In region II, we allow the matterwave to propagate freely in the lattice. The appearance of two wave packets traveling in opposite directions shows that the beam splitter operates as expected. In region III, we apply the mirror shaking function. The matterwave undergoes free propagation in region IV again, and the two wave packets switch directions, demonstrating the functionality of the mirror. Lastly, we apply the recombining protocol in region V. Apart from the main closed diamond-shaped paths, we observe that there are auxiliary paths that are fainter but still clearly evident. They arise due to the imperfection of the components, and also due to the side peaks arising from the third-excited Bloch state.}
\label{fig:diamond}
\end{figure}

We can now cascade together the learned components to make a shaken lattice interferometer. In between the components, we allow the system to propagate freely with the lattice present but with $\phi(t)=0$. This gives a time evolution of the wavefunction in real-space as shown in Fig.~\ref{fig:diamond}, where the state has been initialized to be in the ground state of the lattice multiplied in real-space by a Gaussian envelope with a width of a few lattice sites. The free propagation time is set to $10~\omega_r^{-1}$. During free propagation, we can see the wave-packets propagating at the anticipated $\pm 4\hbar k_L/m$ group velocity. The anticipated features are evident in the observed splitting, reflection and recombination of the matterwave. 

\section{Statistical Analysis} 
\label{sec:analysis}
%%% BAYESIAN %%%
We compute the final momentum distribution on a grid of acceleration values in order to see how useful our device is for inertial sensing.

\begin{figure}
\centering
\includegraphics[width=\columnwidth]{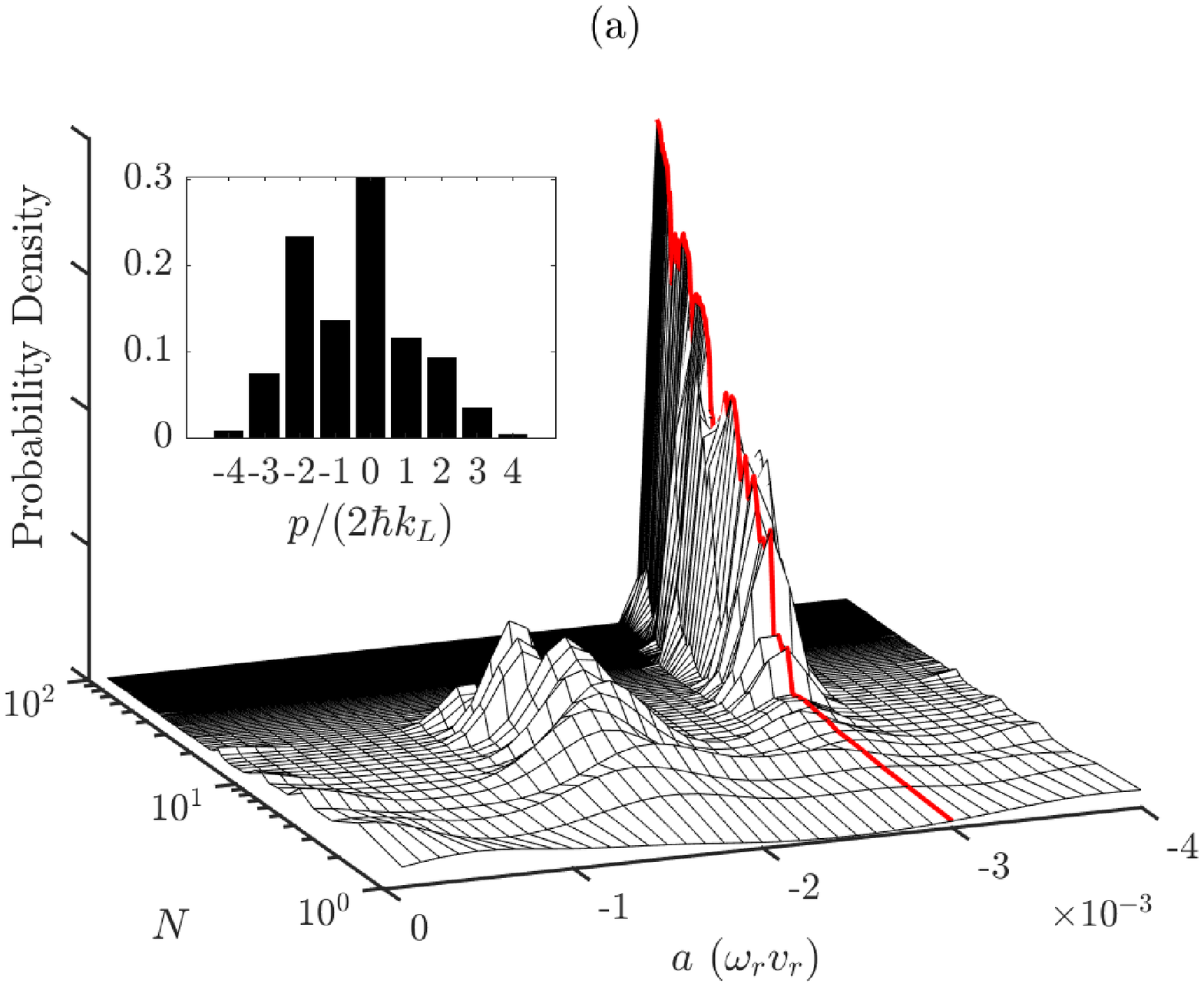}
\includegraphics[width=\columnwidth]{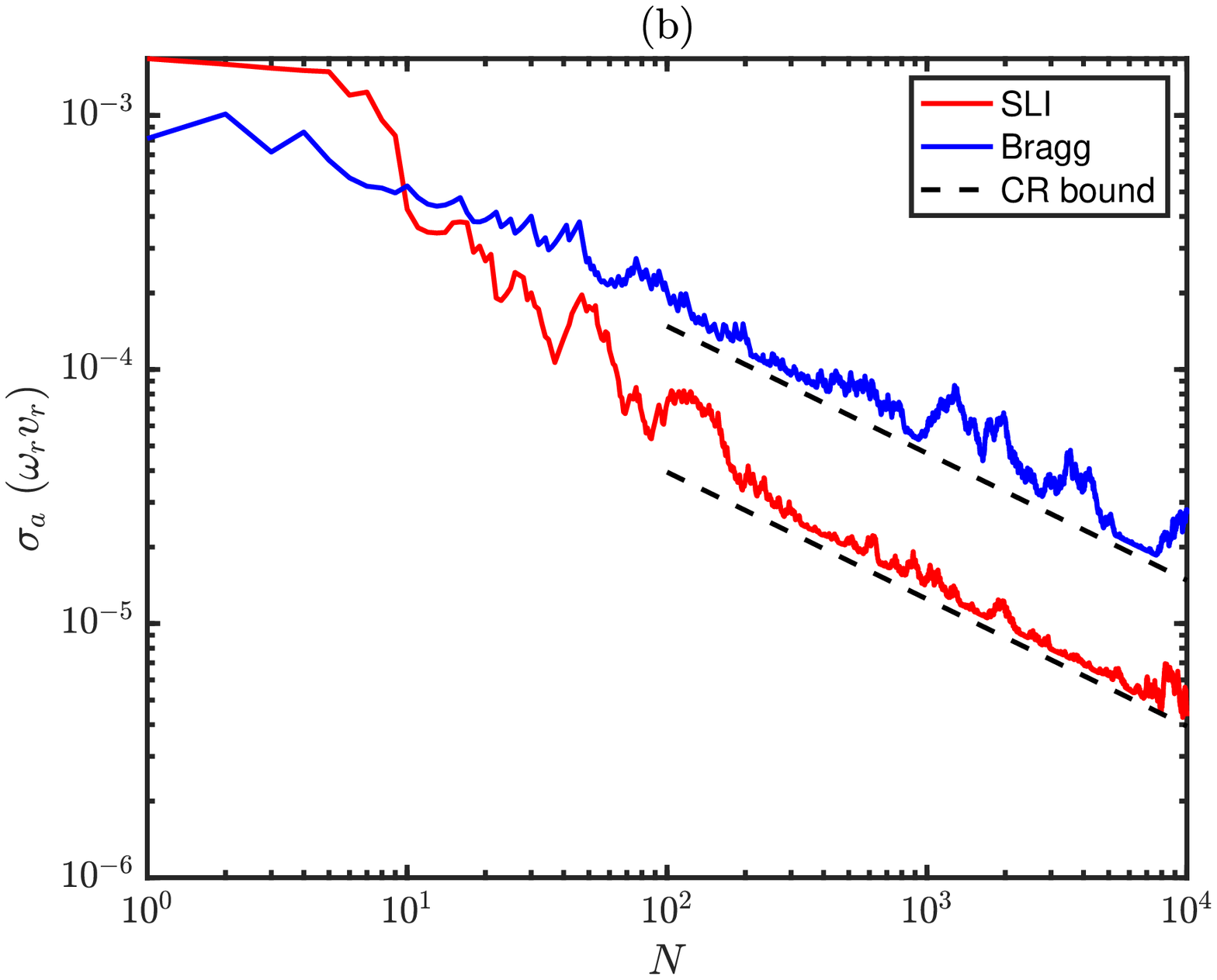}
\caption{(a) Posterior probability density of the acceleration for the first 100  atoms. The true acceleration that we aim to reveal by the measurements is $-3\times 10^{-3}~\omega_r v_r$ (red line), and the measurements are sampled from the momentum distribution at the end of the interferometry sequence, as shown in the inset. (b) Standard deviation of the acceleration estimated using Bayes theorem for up to $10^4$ atoms. We show the results from both the shaken lattice interferometer (SLI) and the Bragg interferometer, and conclude that the shaken lattice interferometer has a higher sensitivity. The standard deviations $\sigma_a$ are roughly inversely proportional to the square-root of the number of measurements $N$, for $N>10^2$. The the black dashed lines are the Cram\'er-Rao (CR) lower bounds, and scale exactly as $1/\sqrt{N}$.}
\label{fig:Bayesian}
\end{figure}
To interpret the outcome of the interferometer, we use Bayes theorem to derive information about the acceleration from the momentum distribution. This approach is always more effective than curve-fitting when there is knowledge of the probability distribution generator, and the difference is particularly notable in situations like this when the distributions are multi-modal.
The probability distribution of the acceleration after measurement of a particle in a particular momentum state, $P(a|p)$, is given by, 
\begin{equation}
    P(a|p) \propto P(p|a)P(a)\,,
    \label{eq:Bayes}
\end{equation}
where $P(a)$ is the prior distribution of the acceleration, and the proportionality is resolved by the normalizing the total probability to unity. The probability for measuring a particular momentum $p$ conditioned on the acceleration~$a$, $P(p|a)$ is directly calculated by propagating the wavefunction in time. 
We combine multiple measurements by iterating Eq.~(\ref{eq:Bayes}) to formulate the conditional probability distribution based on the entire measurement record~\cite{Bayesian}. For a measurement record $\{p_1, p_2, \ldots, p_N\}$, where $N$ is the number of measurements, the probability distribution becomes
\begin{equation}
    P(a|p_1, \ldots, p_N)  \propto P(p_N|a)\ldots P(p_1|a)P(a)\,.
    \label{eq:Bayes_sequence}
\end{equation}
normalized to unity.
Since we assume that each atom is independent from each other, $N$ is the total number of atoms we observe. We combine the distributions for each atom using Eq.~(\ref{eq:Bayes_sequence}) to accumulate the distribution for~$a$ conditioned on the measurement record, and estimate the acceleration by taking the expectation value. 

In Fig.~\ref{fig:Bayesian}(a), we show an example of the parameter estimation process, where the actual acceleration is $-3\times10^{-4}~\omega_rv_r$ ($v_r=\hbar k_L/m$ is the recoil velocity). The corresponding momentum probability distribution for each atom is shown in the inset. What is important about this distribution is that it is almost unique for each acceleration value and therefore acts as a fingerprint. In Figure~\ref{fig:Bayesian}(b), we show the standard deviation of the acceleration as a function of the number of measurements. A typical number of atoms for an ultracold gas experiment may be of order $\sim 10^4$, and with $10^4$ measurements, the estimated acceleration agrees extremely well with the actual value. We observe that the standard deviation approaches the standard quantum limit, which scales as $\sim 1/\sqrt{N}$. This result is consistent with the fact that we are not considering atom interactions in the system, so each atom may be thought of as making an independent measurement.  

We demonstrate that reinforcement learning provides additional capability over traditional experimental techniques by comparing the sensitivity of the shaken lattice interferometer with the conventional Bragg interferometer. The Bragg interferometry sequence is shown in Fig.~\ref{fig:Intro}(b). 
%During the $\pi/2-$ and $\pi-$ pulses, the lattice is turned on with a fixed amplitude and no phase modulation, and during free propagation the lattice is turned off. 
We use the Bayesian parameter estimation result with the Bragg interferometer as a baseline for benchmarking the shaken lattice interferometer. To make a fair comparison, we consider the case where the Bragg interferometry sequence takes the same time as the shaken lattice interferometer. The protocol found by our reinforcement learning algorithm generates a momentum splitting of $8\hbar k_L$ (between $-4\hbar k_L$ and $+4\hbar k_L$) at the beam splitter, 4 times larger than the $2\hbar k_L$ (between $-1\hbar k_L$ and $+1\hbar k_L$) splitting from Bragg diffraction, and therefore this results in higher sensitivity. One may argue that it is possible to generate larger momentum splitting with Bragg diffraction, and while this is true, higher-order transitions typically require much more time than than the short pulses applied here. For these reasons, we have observed from our simulations that the standard deviation of the shaken lattice interferometer can be approximately 4 times lower than that of the corresponding Bragg interferometer given the  time constraint, implying that we have realized an approximate factor of 4 in sensitivity gain. 
The results are shown in Fig.~\ref{fig:Bayesian}(b).

We verify that the standard deviations $\sigma_a$ for both the shaken lattice interferometer and the Bragg interferometer are close to the limits set by the Cram\'er-Rao bound, which can be calculated from the classical Fisher information as,
\begin{eqnarray}
    \sigma_a &\geq& 1/\sqrt{NI_1(a)}, \nonumber\\
    I_1(a) &=& \sum_p \frac{1}{P(p|a)}\left[\pd{P(p|a)}{a}\right]^2 \,,
\end{eqnarray}
where $I_1(a)$ is the Fisher information for one independent measurement.
The Cram\'er-Rao bound for the shaken lattice interferometer is determined by numerically calculating the classical Fisher information using the pre-calculated probability $P(p|a)$, and the Cram\'er-Rao bound for the Bragg interferometer is determined by the analytical form $\sigma_a \geq (2k_LT^2)^{-1}N^{-1/2}$, where $T$ is the free propagation time.

\section{Conclusion}
In summary, we have demonstrated a machine learning methodology to control a complex quantum system for the purpose of performing a quantum metrology task.  Specifically, we demonstrated how to utilize reinforcement learning for the design of a lattice-based matterwave interferometer. We showed that by shaking the lattice with protocols derived from deep Q-learning, matterwave analogs of optical components are realized. We showed that these can be concatenated together to build a high-precision interferometric device. The multipath interferometer that was constructed in this way is capable of the measurement of acceleration with higher sensitivity than that achieved by a conventional matterwave interferometer.

While we have assumed that atom interactions are negligible, they could potentially bring more quantum advantage if harnessed appropriately. For example, if we could discover a protocol for shaking the lattice to generate momentum-squeezed states, the sensitivity could potentially be higher than the standard quantum limit that is generated by the statistical averaging of independent atoms. In fact, the quantum entanglement in such squeezed states could allow the resulting interferometer to approach the ultimate limit to sensitivity set by the Heisenberg uncertainty relation between number and phase. We have focused on demonstrating a specific solution to the problem of designing a Mach-Zehnder interferometer, but our main outcome is actually to illustrate an effective design methodology. In the future, this approach could potentially lead to the design of alternate protocols that construct completely new types of metrological devices with performance level surpassing conventional experimental methods.

%%%%%%

\begin{acknowledgments}
The authors acknowledge helpful discussions with Dana Anderson, Marco Nicotra, Claire Monteleoni, Haonan Liu, and Simon J\"ager.
This work was supported by NSF Grant No. 1936303, NSF QLCI Award No. OMA-2016244, NSF PFC Grant No. 1734006; and the DARPA and ARO Grant No. W911NF-16-1-0576.
\end{acknowledgments}

\appendix

\section{Training the deep Q-network} \label{app:train}
The objective of deep Q-learning is to train the deep Q-network so that it represents the return defined in Eq.~(\ref{eq:series}) well and therefore satisfies the Bellman optimality equation,
\begin{equation}
    Q(s,a) = r(s') + \gamma \max_{a'}Q(s',a') \,.
    \label{eq:Bellman}
\end{equation}
We train the deep Q-network using the double deep Q-learning algorithm \cite{doubleDQN}. In this algorithm, an extra network called the target network is included to estimate the future return [Eq.~(\ref{eq:future})]. The reason to do it this way is that Eq.~(\ref{eq:Bellman}) involves a self-consistency element. If during the training stage, the Q-values on both sides of the equation are determined by the same Q-network, then updating the Q-network will simultaneously change both the Q-value $Q(s,a)$ and the return $Y(s')$ that the Q-value is supposed to converge to. This creates a feedback cycle that can make the learning process extremely unstable, and in fact the Q-values may never converge. In order to mitigate this problem, we employ a Q-network, $Q_\vectorsym{\theta}$, just as described, from which the optimal action is chosen by the agent through their policy. However, we then update the return used to optimize the Q-network with a Q-value corresponding to this action but obtained from a second neural network, the target network, $Q_{\vectorsym{\theta}'}$. The target network is only updated more slowly through a weighted average. This avoids the unstable feedback since the Q-network and target network are weakly correlated at early stages of the learning process \cite{DQNNature}. 
Furthermore, this method also resolves issues with overoptimism in the standard deep Q-learning algorithm \cite{doubleDQN}. A sketch of the algorithm is presented in Algorithm~\ref{alg:dDQN}.
\begin{algorithm}%[H]
\label{alg:dDQN}
\SetAlgoLined
%\KwResult{}
 Initialize Q-network $Q_\vectorsym{\theta}$ and target network $Q_{\vectorsym{\theta}'}\leftarrow Q_\vectorsym{\theta}$\;
 \For{trajectory=1:episodes}{
     $s\leftarrow s_0$\;
     \While{d=0}{
          \eIf{$rand() > \epsilon$}{
           $a \leftarrow \arg\max_a Q_\vectorsym{\theta}(s,a)$\;
           }{
           $a \leftarrow rand(\mathcal{A})$\;
           }
           $[s',r,d]\leftarrow environment(s,a)$\;
           Store $(s,a,s',r,d)$ in replay buffer\;
           Sample from replay buffer: $\{(s_i,a_i,s'_i,r_i,d_i)|\,i=1,...,B\}$\;
           $\mathbb{L} = \frac{1}{B}\sum_i\left[Q_\vectorsym{\theta}(s_i,a_i) - Y_i\right]^2$,
            $Y_i = r_i+\gamma\, Q_{\vectorsym{\theta}'}(s_i',\arg\max_{a'}Q_\vectorsym{\theta}(s_i', a_i'))$\;
           Update $Q_\vectorsym{\theta}$: $\vectorsym{\theta} \leftarrow \vectorsym{\theta} - \alpha \nabla_\vectorsym{\theta}\mathbb{L}$\;
           Update $Q_{\vectorsym{\theta}'}$: $\vectorsym{\theta}' \leftarrow \tau\vectorsym{\theta}' + (1-\tau) \vectorsym{\theta}$\;
           $s\leftarrow s'$\;
    }
 }
 \caption{Double Deep Q-Learning}
\end{algorithm}

The algorithm is described as follows. First of all, we initialize the weights and biases in the Q-network randomly, and initialize the target network such that its network parameters are the same as those of the Q-network. 
Then, we iterate over a large number of episodes. In each episode, a trajectory of consecutive states and actions is generated. A trajectory starts with an initial state $s_0$ and ends when we reach a terminal state or a threshold number of steps.

In each step, we choose the action that maximizes the Q-value for the current state most of the time, but with probability $\epsilon$ we choose a random action. In practice, $\epsilon$ decays from 1 to 0.01 over time, since exploitation becomes more important than exploration as the learning process proceeds. The action is then fed into the environment, and the environment outputs the subsequent state $s'$, a reward $r$, and a Boolean $d$ that indicates whether the trajectory terminates at this step. A tuple $(s, a, s', r, d)$, which represents the experience we get from taking this step, is pushed into the replay buffer.

To minimize the error of $Q_\vectorsym{\theta}(s, a)$ for representing the return, we calculate a loss function defined as the squared difference between $Q_\vectorsym{\theta}(s, a)$ and $Y(s')$. 
The future return in $Y(s')$ [Eq.~(\ref{eq:future})] is evaluated by the output value of the target network for the next optimal action. The next optimal action is determined by the action that maximizes the output values of the Q-network, given the next state $s'$ as input. Since we need the Q-values to do well for all possible pairs of $(s, a)$, we draw $B$ (batch size) samples from the replay buffer, where each sample is labelled by an index $i$,
and calculate their average loss. 

The parameters in the Q-network are updated through gradient descent in the direction of decreasing average loss. The parameters in the target network are updated by taking the weighted average of the previous parameters from the target network and the updated parameters from the Q-network. 

\section{Network topology} \label{app:network}
The Q-network we use includes an input layer, one hidden layer, and an output layer. 
Each layer is connected to the next one by a linear function followed by a ReLU (Rectified Linear Unit) activation function, which represents the function $\max(x, 0)$ with the argument $x$ being a real number. The relations between layers are explicitly written out as follows,
\begin{eqnarray}
    \vectorsym{h} &=& \mathrm{ReLU}(\mathbf{W}_1\vectorsym{x}+\vectorsym{b}_1)\\
    \vectorsym{y} &=& \mathrm{ReLU}(\mathbf{W}_2\vectorsym{h}+\vectorsym{b}_2)\,,
\end{eqnarray}
where $\vectorsym{h}$ denotes the hidden nodes, $\vectorsym{x}$ denotes the input nodes, $\vectorsym{y}$ denotes the output nodes, $\mathbf{W}_1,\ \mathbf{W}_2$ are the matrices of weights, and $\vectorsym{b}_1,\ \vectorsym{b}_2$ are vectors of biases.
\begin{figure}
\centering
\includegraphics[width=0.4\columnwidth]{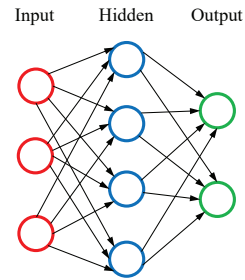}
\caption{An example topology of the network that we used in the deep Q-algorithm.}
\end{figure}

\section{Reinforcement learning for splitting} \label{app:RL_split}
 Here we present how we formulate the control for splitting as a reinforcement learning problem. First of all, we define the input state to be a vector of the populations in the momentum eigenstates, 
 \begin{equation}
     \{|2n\hbar k_L\rangle\mid n=-3,-2,-1,0,1,2,3\}\,.
     \label{eq:states1}
 \end{equation}
 The momentum states that are neglected here are barely populated the whole time. The actions are chosen from a set of discrete phase values, 
 \begin{equation}
      \{n\pi/4\mid n=-4,-2,-1,0,2\}\,.
      \label{eq:actions1}
 \end{equation}

 The reward is a function of fidelity, defined as
 \begin{eqnarray}
    r &=& \begin{cases}
    0 & \quad\text{if }d=0\\
    \frac{\mathcal{F}}{1-\mathcal{F}} & \quad\text{if }d=1
    \end{cases}
    \,,\label{eq:reward}\\ 
    \mathcal{F} &=& |\braket{\psi_\mathrm{target}}{\psi(t)}|^2\,,
\end{eqnarray}
where $d=1$ when the maximum terminal time is reached or when the fidelity is higher than the threshold $0.95$, and $d=0$ otherwise. 
The Q-network we use for this task has an input dimension of 7 and an output dimension of 6, as a result of (\ref{eq:states1}) and (\ref{eq:actions1}). The neural network is implemented in PyTorch. The values for the hyperparameters used for training this Q-network are listed in Table \ref{tab:hyper1}.
\begin{table}[ht]
  \centering
  %\footnotesize
  \begin{tabular}{ll}
    \toprule
    Hyperparameters & Values \\
    \midrule
    $\gamma$ & 0.999\\
    $\tau$ & 0.999\\
    $\alpha$ & 0.001\\
    $episodes$ & 20,000\\
    $\epsilon$ decay & 0.0001\\
    Hidden size & 98\\
    Batch size & 64\\
    Optimizer & Adam\\
    \bottomrule
  \end{tabular}
  \caption{Choice of hyperparameters for the splitting task. The exploration probability $\epsilon$ decays linearly for each episode at the rate represented by $\epsilon$ decay. The optimizer we use, `Adam' (which stands for adaptive momentum estimation), is a method for efficient stochastic gradient descent \cite{Adam}.}
  \label{tab:hyper1}
\end{table}

\section{Reinforcement learning for the mirror} \label{app:RL_mirror}
The reflection operation is defined by a target unitary operator that satisfies Eq.~(\ref{eq:mirror}). Our goal is to control the phase as a function of time, such that the unitary operator 
\begin{equation}
    \hat{U}(t) = \mathcal{T}\left\{ \exp \left[\int_0^t-i\hat{H}(\phi(t'))dt'/\hbar\right]\right\}\,,
\end{equation}
reaches the target operator at the terminal time. Here $\mathcal{T}$ represents the time-ordering operator and $\hat{H}(\phi)$ is defined in Eq.~(\ref{eq:Hamiltonian}).
To formulate the control problem as a reinforcement learning problem, we assign the state to be the matrix elements of the unitary operator. 
Since before the wavefunction enters the reflection stage, most of its population is in the $\ket{\pm 4\hbar k_L}$ subspace, we only keep track of the operation in this subspace. 
Because of this,  we reduce the input state to the modulus square of the matrix elements for only 
\begin{equation}
\{\ket{2n\hbar k_L}\bra{\pm4\hbar k_L}|n=-3,-2,-1,0,1,2,3\}.
\label{eq:states2}
\end{equation}
The state is chosen so that it forms an experimental observable. This state can be obtained by initializing the quantum state to one of the $\ket{\pm4\hbar k_L}$ states, applying the control and measuring the populations for all the momentum states through time-of-flight imaging.

As discussed in the main text, we found that the frequency of the shaking function is important for the performance of the mirror. Simply by modulating
at the characteristic frequency, $12 \omega_r$, which corresponds to the energy difference between $\ket{4\hbar k_L}$ and $\ket{2\hbar k_L}$, with a fixed amplitude for a duration of $\sim3\omega_r^{-1}$, the channel fidelity can reach as high as $0.8$. We take advantage of this knowledge, and try to improve the fidelity upon the fixed-amplitude modulation. We choose our actions to be the amplitude of the sinusoidal modulation to the phase, so the resulting phase is $\phi(t) = \mathrm{Amp}(t)\times \sin(12\omega_r t)$. The amplitudes are chosen from a discrete set of values, 
\begin{equation}
\mathrm{Amp}=\{0.4,\,0.6,\,0.8,\,1.0,\,1.2\}.
\label{eq:actions2}
\end{equation}
The time interval between each decision point is $\pi/12 \approx0.26~\omega_r^{-1}$, and in each interval the amplitude is held constant.

The reward function is the same as Eq. (\ref{eq:reward}), except that the fidelity $\mathcal{F}$ is now defined with the channel fidelity in the relevant subspace [see Eq.~(\ref{eq:channel-fidelity})].

With the states, actions, rewards specifically designed for reflection, we use the double deep Q-learning algorithm to learn the strategy for controlling the amplitude of the sinusoidal modulation to the phase.  
The deep Q-network used for learning the mirror has an input size of 14, and an output size of 5, as a result of (\ref{eq:states2}) and (\ref{eq:actions2}). The hyperparameters are listed in Table~\ref{tab:hyper2}.
\begin{table}[ht]
  \centering
  %\footnotesize
  \begin{tabular}{ll}
    \toprule
    Hyperparameters & Values \\
    \midrule
    $\gamma$ & 0.99\\
    $\tau$ & 0.99\\
    $\alpha$ & 0.001\\
    $\epsilon$ decay & 0.0005\\
    $episodes$ & 8,000\\
    Hidden size & 128\\
    Batch size & 32\\
    Optimizer & Adam\\
    \bottomrule
  \end{tabular}
  \caption{Choice of hyperparameters for the reflection task.}
  \label{tab:hyper2}
\end{table}

%\bibliography{ref.bib}
%

\end{document}